\documentclass{mn2e}
\usepackage{natbib}
\begin{document}
\title[A simple derivation and interpretation of the third integral in
Stellar Dynamics]
  {A simple derivation and interpretation of the third integral in
Stellar Dynamics}
\author[D Lynden-Bell]
  {D Lynden-Bell\\
The Observatories of the Carnegie Institution,
813 Santa Barbara St., Pasadena, CA., U.S.A.\\
  The Institute of Astronomy, The Observatories, Madingley Road,
  Cambridge, CB3 0HA, U.K. (permanent address), \\
and Clare College, Cambridge.}
\date{Accepted . Received }
\pagerange{\pageref{firstpage}--\pageref{lastpage}}
\pubyear{}
\label{firstpage}
\maketitle
\begin{abstract}
Starting from the problem of two fixed centres we find a simple
derivation of its third integral in terms of the scalar product of the
angular momenta about the two fixed centres.  This is then generalised
to find the general form of the potential in which an exact third
integral exists.
\end{abstract}
\section{Introduction}
When deriving or using the third integral of the motion, i.e., that
other than energy or angular momentum about the axis, I have often
been asked for its physical meaning.  The new derivation given here
may not completely answer that question, but the realisation that the
kinetic part is the dot product of the angular momenta about the two
centres, or two foci, goes somewhat further towards an answer than
anything I have seen previously.  

This paper gives what I hope is a simpler, cleaner derivation of the
integral, in elementary terms that do not require spheroidal
coordinates or the separation of the Hamilton--Jacobi equation.
Earlier treatments requiring such sophistications are found in
\citet{lyn}, \citet{ed15}, \citet{dez}, \citet{sta}  
and \citet{kuz}.  The two dimensional problem is discussed in
\citet{whi}.  \citet{con} showed that in many galactic
potentials of non separable form, a regular third integral exists for
most orbits.  
\section{Preliminaries - The One Centre Problem}
Hamilton's beautiful treatment of this problem is not widely
enough taught \citep{ham}.  As we need some of the steps later we give his
treatment here.  The equation of motion is 
\begin{equation}
\ddot{{\bf r}} = - \mu \hat{{\bf r}}/r^2 \ , \label{eq1} 
\end{equation}
where $\mu  =Gm$ and $\hat{{\bf r}}$ is the unit vector.  Cross
multiplying by ${\bf r}$ and writing ${\bf h}= {\bf r} \times
\dot{{\bf r}}$ we find 
\begin{displaymath}
\dot{{\bf h}} = {\bf r} \times \ddot{{\bf r}} = 0 \ ,
\end{displaymath}
so ${\bf h}$ is constant.  Cross multiplying  (\ref{eq1}) by ${\bf h}$ 
\begin{equation}
d \left ({\bf h} \times \dot{{\bf r}} \right )/dt  = 
-\mu \left (\hat{{\bf r}} \times \dot{{\bf r}}/r\right ) \times
\hat{{\bf r}}  =  -\mu d \hat{{\bf r}}/dt \ ,\label{eq2}
\end{equation}
the last identity may readily be demonstrated by dull algebra but the
physical understanding is more interesting.  In a small interval
$\delta t,\ \hat{{\bf r}} \times \delta {\bf r}/r$ is perpendicular to
the movement $\delta {\bf r}$ and to $\hat{{\bf r}}$ and gives the
angle moved about the centre.  Hence $\hat{{\bf r}} \times \dot{{\bf
    r}}/r$ is the angular velocity of the radius vector which we call
$\mbox{\boldmath$\omega$}$.  Hence $(\hat{\bf r} \times \dot{\bf r}/r)
\times \hat{\bf r} = \mbox{\boldmath$\omega$}\times \hat{{\bf r}}$
where $\mbox{\boldmath$\omega$}$ is the angular velocity of the unit
vector $\hat{{\bf r}}$.  However, the change in a unit vector is just
$\mbox{\boldmath$\omega$} \times \hat{{\bf r}}$ because it only
changes by rotation since its length is fixed so $d\hat{{\bf r}}/dt =
\mbox{\boldmath$\omega$} \times \hat{{\bf r}}$.

Equation (\ref{eq2}) is readily integrated to give 
\begin{equation}
\dot{\bf r} \times {\bf h} = \mu (\hat{\bf r} + {\bf e})\ ,
\label{eq3}
\end{equation}
where ${\bf e}$ is a constant of integration.  Since both $\hat{\bf
  r}$ and $\hat{\bf r} \times {\bf h}$ are perpendicular to ${\bf h}$
  we find ${\bf e}\cdot {\bf h} =  0 $ .  Take the scalar product of
  (\ref{eq3}) with $\hat{{\bf r}}/\mu$ then, on use of 
\begin{displaymath}
\hat{\bf r}\cdot (\dot{{\bf r}} \times {\bf h}) = ({\bf r}\times \dot{\bf
  r})\cdot {\bf h}/r = h^2/r
\end{displaymath}  
we find 
\begin{equation}
\ell/r = 1+{\bf e}\cdot \hat{\bf r}\ , 
\label{eq4}
\end{equation}
where $\ell = h^2/\mu$.  This is the equation of a general conic where
${\bf e}\cdot \hat{\bf r} = e \cos \phi$ with $\phi$ measured from
pericentre and $e$ the eccentricity.  So we now identify Hamilton's
vector ${\bf e}$ as the vector that points to the pericentre and whose
length is the eccentricity.  We note in passing the useful expression
for the velocity found from (\ref{eq3}), 
\begin{equation}
\dot{\bf r } = \mu h^{-2} {\bf h} \times \left (\hat{\bf r} + {\bf
    e}\right ) \ . \label{eq5}
\end{equation} 
Its square may be used to derive the energy integral
\section{The Third Integral for the two centre problem}
We measure ${\bf r_1}$ from the first centre of attraction at $-{\bf
  a}$ and ${\bf r_2}$ from the second at $+{\bf a}$ so ${\bf r_1} =
  {\bf r_2} + 2{\bf a}$.  The equation of motion is 
\begin{equation}
\ddot{{\bf r_1}} = \ddot{{\bf r}}_2 = - \mu_1 \hat{{\bf r}}_1/r^2_1 - \mu_2
    \hat{\bf r}_2/r^2_2\ . \label{eq6}
\end{equation}
The energy and the component of angular momentum along the line of
centres are, of course, conserved.  We are interested in an
independent third integral.  Cross multiplying (\ref{eq6}) by ${\bf
  r_1}$ we find
\begin{displaymath}
\dot{\bf h}_1 = 
d \left ( {\bf r}_1 \times \dot{\bf r_1}\right )/dt
= - \mu_2 {\bf r}_1 \times \hat{\bf r}_2/r^2_2 
= - \mu_2 2{\bf a} \times \hat{\bf r}_2/r^2_2\ ,
\end{displaymath}
taking the scalar product with ${\bf h}_2$ we find {\it c.f.} (\ref{eq2}), 
\begin{displaymath}
{\bf h}_2 \cdot \dot{\bf h}_1 = 
-\mu_2 2{\bf a}\cdot 
\left [
\hat{{\bf r}}_2 \times 
\left (\hat{{\bf r}}_2 \times \dot{{\bf r}}_2/r_2 
\right )
\right ]
= \mu_2 2{\bf a} \cdot d\hat{\bf r}_2/dt \ .
\end{displaymath}
Adding a half of this to the conjugate expression with ${\bf r}_2$ and
${\bf r}_1$ exchanged and $-{\bf a}$ for ${\bf a}$ we find
\begin{equation}
d\left ({\scriptstyle{1 \over 2}} {\bf h}_1 \cdot {\bf h}_2
\right )/dt = d 
\left [{\bf a}\cdot 
\left (\mu_2 \hat{\bf r}_2 - \mu_1 \hat{\bf r}_1 
\right )
\right ]/dt \ . \label{eq7}
\end{equation}
So
\begin{equation}
I_3 = {\scriptstyle{1 \over 2}}  {\bf h}_1 \cdot  {\bf h}_2 +
\left (\mu_1 \hat{\bf r}_1 -\mu_2 \hat{\bf r}_2 
\right )
\cdot {\bf a} = {\rm constant} \ , \label{eq8}
\end{equation}
which is the third integral for this problem.  Using 
\begin{displaymath}
\widetilde{r} =
{\scriptstyle{1 \over 2}} (r_1 + r_2)\ \ ; \ \ 
a \widetilde{\mu} = (r_1 - r_2)/2
\end{displaymath} 
as coordinates the three integrals may now be used to solve for the
orbits, but our purpose here is to generalise this beautifully simple
derivation of (\ref{eq8}) to more general problems.  When $\mu_2=0$,
${\bf h}_2 = {\bf h}_1 - 2{\bf a} \times \dot{r}_1$ and the integral
reduces to a combination of ${\bf h}_1$ and ${\bf e}$.
\section{The exact third integral more generally}
We write our axially symmetrical potential $\psi$ in the form
$\psi(r_1,\ r_2)$.  The equation of motion is
\begin{displaymath}
\ddot{{\bf r}}_1 = \ddot{{\bf r}}_2 = \mbox{\boldmath$\nabla$} \psi =
\partial \psi/\partial r_1 \hat{\bf r}_1 + \partial \psi/\partial r_2
\hat{\bf r}_2\ .
\end{displaymath}
Performing the same steps as led to (\ref{eq7}) we find in its place 
\begin{equation}
d\left({\scriptstyle{1\over 2}} {\bf h}_1 \cdot {\bf h}_2\right )/dt
= - r^2_2 \partial \psi/\partial r_2 
{\bf a}\cdot d \hat{\bf r}_2/dt +
r^2_1 d\psi/dr_1 
{\bf a}\cdot d\hat{\bf r}_1/dt\ , \label{eq9}
\end{equation}
which reduces to our former result when $\psi = \mu_1/r_1 +
\mu_2/r_2$.  

To proceed we need expressions for ${\bf a}\cdot \hat{\bf r}_2$, etc.
These come from squaring the expression for ${\bf r}_1$ in terms of
${\bf a}$ \linebreak and ${\bf r}_2$.
\begin{displaymath}
r^2_1 =r^2_2 + 4a^2 + 4{\bf a}\cdot \hat{\bf r}_2 r_2
\end{displaymath}
so
\begin{equation}
r^2_2 d \left ({\bf a} \cdot \hat{r}_2 \right )/dt 
= {\scriptstyle{1\over 2}} r_1 r_2 \dot{r}_1 -
\left [{\scriptstyle{1\over 4}}
\left (r^2_1 + r^2_2 
\right )
-a^2 \right ] \dot{r}_2 \ . \label{eq10}
\end{equation}
Collecting terms in $\dot{r}_1$ and in $\dot{r}_2$ in (\ref{eq9}) we
have using (\ref{eq10}) and its conjugate with $r_1\longleftrightarrow
r_2$ and $a \longrightarrow -a$ 
\begin{eqnarray}
d\left ({\scriptstyle{1\over 2}} {\bf h}_1 \cdot {\bf h}_2 \right )/dt
= 
\left \{ - {\scriptstyle{1\over 2}} r_1 r_2 \partial \psi/\partial
  r_2) \right. \\
\left. + \left [ {\scriptstyle{1\over 4}}
\left (r^2_1 + r^2_2 \right )
-a^2 \right ] d\psi/dr_1 \right \} 
\dot{r}_1 + {\rm conjugate} \ . \label{eq11}\nonumber
\end{eqnarray}\nonumber

This last expression may be rewritten
\[
\left \{ 
\begin{array}{l}
\vspace{10pt}{-{\scriptstyle{1\over 2}}} \partial/\partial r_2 \left (
r_1 r_2 \psi \right ) +\vspace{10pt} \\
+ {\scriptstyle{1\over 4}} \partial/\partial r_1 \left [\left (r^2_1 +r^2_2 -4a^2 \right )\psi \right ] 
\end{array} 
\right \} \dot{r}_1 + {\rm conjugate}\ .
\]
The second term in the curly bracket gives rise to an expression
which, taken with its conjugate is a total time derivative.  

We now ask for the condition on $\psi$ that the remainder plus its
conjugate be a perfect derivative.  For this to be true it must be of
the form $\partial \Lambda/\partial r_1\ \dot{r}_1 +
\partial\Lambda/\partial r_2\ \dot{r}_2$ so $\partial/\partial r_2$ 
of the first term must be $\partial/\partial r_1$ of its conjugate. 

Writing $\chi = r_1 r_2 \psi$ our condition
reduces to
\begin{displaymath}
\partial^2\chi/\partial r^2_1 - 
\partial^2\chi/\partial r^2_2 = 0\ ,
\end{displaymath}
but this is the wave equation and its general solution is 
\begin{displaymath}
\chi = \zeta (r_1 + r_2) - \eta (r_1 - r_2)\ , 
\end{displaymath}
where $\zeta$ and $\eta$ are arbitrary functions of those arguments.
Thus the potential $\psi$ must take the form 
\begin{displaymath}
\psi = \left [\zeta (r_1 + r_2) - \eta (r_1-r_2) \right ]/(r_1r_2) \ .
\end{displaymath}
With this expression for $\psi$, (\ref{eq11}) takes the form
\begin{eqnarray*}
{d \over dt}
\left ( {\scriptstyle{1\over 2}} {\bf h}_1 \cdot {\bf h}_2 \right )
=\\ 
= {\partial \over \partial r_1}
\left \{
\left [
{\scriptstyle{1\over 4}} 
\left (r^2_1 + r^2_2 \right )
-a^2 \right ]
\psi - {\scriptstyle{1\over 2}} \left (\zeta + \eta \right )
\right \} \dot{r}_1 + \\ 
+ {\rm conjugate}, 
\end{eqnarray*}
so
\begin{eqnarray*}
I_3 & =  & {\scriptstyle{1\over 2}} {\bf h}_1 \cdot {\bf h}_2 -
\left [ {\scriptstyle{1\over 4}}
\left (r^2_1 + r^2_2 \right )
-a^2 \right ]
{\zeta - \eta \over r_1 r_2} + 
 {\scriptstyle{1\over 2}} 
\left (\zeta + \eta \right ) = {\rm const} \\
& = & {\scriptstyle{1\over 2}}  
{\bf h}_1 \cdot {\bf h}_2
- \\
& & {1 \over r_1r_2}
\left \{ \left [ \left (
{r_1 - r_2 \over 2} \right )^2
- a^2 \right ]
\zeta - 
\left [\left (
{r_1 + r_2 \over 2} \right )^2
-a^2 \right ]
\eta \right \}\ . 
\end{eqnarray*}
The final term reduces with suitable definitions to the 
$-(\mu \zeta - \lambda \eta)/(\lambda - \mu)$ form one finds in
standard treatments \citep{lyn}.

When the coordinates are oblate ${\bf a} $ is imaginary $i\vert {\bf
  a} |$ so ${\bf h}_1$ and ${\bf h}_2$ are angular momenta $({\bf r} +
  i \vert {\bf a} \vert ) \times \dot{{\bf r}}$, etc, about the
  imaginary points ${\bf r} = \pm i \vert {\bf a}\vert $.  However the
  product of these angular momenta is still real.  The expressions for
  the integral are unchanged 
except that when the $r_1$ and $r_2$ are interpreted as the greatest
and least distances of a general point from the ring $z=0,\ {r}=
\vert a \vert$ then the $-a^2$ terms in the final expression of $I_3$
should be omitted.  

I am unable to extend this derivation into the relativistic r\'{e}gime
while  maintaining its simplicity.  \citet{car68} has shown that a
third integral exists for a charged particle moving in the Kerr-Newman
metric.  When Newton's $G$ is set equal to zero that metric leaves
behind the most interesting electromagnetic field with a net charge
$q_1$ (discussed in \citet{lyn01}).
\begin{displaymath}
{\bf E} + i{\bf B} = - \nabla\left[q_1/\sqrt{({\bf r}-i{\bf a})\cdot
    ({\bf r}-i{\bf a})}\right]\ \ \ ; \ \ \ {\bf a} = (0,0,a)\ .
\end{displaymath}
Third integrals in this and other electromagnetic fields in the
relativistic r\'{e}gime are related to the classical third integral
discussed above.  For a demonstration of this relationship and a
derivation of the forms of electromagnetic fields in which they exist
see \citet{lyn00}.  Here we merely quote the form the third integral
takes for the above electromagnetic field with the orbiting particle
of mass $m$ and charge $q$. 
$\mbox{\boldmath$\hat\phi$}$ is the unit toroidal vector, $\Phi$ and
${\bf A} = A \mbox{\boldmath$\hat\phi$}$ the electromagnetic scalar
and vector potentials.

\begin{eqnarray*}
I& = & {\scriptscriptstyle{1\over2}} {\bf h}_1 \cdot {\bf h}_2 + q\mu
\zeta_3(\lambda)/(\lambda -\mu) \\
\parbox{1cm}{where}\\
{\bf h}_1 & =&  ({\bf r}-i{\bf a}) \times {\bf p} \ \ ; \ \ {\bf h}_2 =
({\bf r} + i{\bf a}) \times {\bf p}\\
\zeta_3(\lambda) & = & q_1 \sqrt{\lambda -a^2}\ \lambda^{-1} c^{-2}
\left (\epsilon \lambda - {\scriptscriptstyle{1\over 2}} qq_1 \sqrt{\lambda
  -a^2}-ach\right)\\
{\bf p} & = & m{\bf v} /\sqrt{1-V^2/c^2} + q A
\mbox{\boldmath$\hat\phi$}/c \ \ ; \ \ {\bf v}=\dot{\bf{r}} \\
\epsilon & = & mc^2/\sqrt{1-V^2/c^2} + q \Phi\\
h & = & mR^2 \dot{\phi}/\sqrt{1-V^2/c^2} + qRA/c\\
\Phi & = & q_1 \sqrt{\lambda -a^2}/(\lambda - \mu) \\
A/R & = & a^2\lambda^{-1}\Phi\\
2\lambda & = & r^2+a^2+\vert \left({\bf r}-i{\bf a}\right )^2\vert\\
& & \hspace{.175cm}\parbox[t]{7cm}{(constant on oblate confocal spheroids)}\\
2\mu &  = & r^2 + a^2 - \vert \left({\bf r}-i{\bf a}\right )^2\vert \\
& & \hspace{.175cm}\parbox[t]{6cm}{(constant on confocal hyperboloids of one sheet)}
\end{eqnarray*}
$I,\ h$ and $\epsilon$ are the integrals,
$R^2=x^2+y^2=\lambda\mu/a^2$, and the $\vert\ \ \vert$ sign is used in
the sense of complex numbers. 
\section*{References}
\begin{itemize}
\bibitem[\protect\citeauthoryear{Carter}{1968}]{car68}
Carter, B., 1968, Comm.Math.Phys, 10, 280
\bibitem[\protect\citeauthoryear{Contopoulos}{1960}]{con}
Contopoulos, G., 1960, Z.Astrophys., 49, 273
\bibitem[\protect\citeauthoryear{de Zeeuw}{1985abc}]{dez}
de Zeeuw, P.T., 1985, MNRAS, 215, 731
\bibitem[\protect\citeauthoryear{de Zeeuw}{1985}]{dez2}
de Zeeuw, P.T.,  1985, MNRAS, 216, 273 \& 599
\bibitem[\protect\citeauthoryear{Eddington}{1915}]{ed15}
Eddington, A.S., 1915, MNRAS, 76, 37
\bibitem[\protect\citeauthoryear{Hamilton}{1847}]{ham}
Hamilton, W.R., 1847, Proc.R.Irish.Acad, 3, Appendix p36
\bibitem[\protect\citeauthoryear{Kuzmin}{1956}]{kuz}
Kuzmin, G.G., 1956, Astr.Zh., 33, 27
\bibitem[\protect\citeauthoryear{Lynden-Bell}{1962}]{lyn}
Lynden-Bell, D., 1962, MNRAS, 124, 95
\bibitem[\protect\citeauthoryear{Lynden-Bell}{2000}]{lyn00}
Lynden-Bell, D., 2000, MNRAS, 312, 301
\bibitem[\protect\citeauthoryear{Lynden-Bell}{2001}]{lyn01}
Lynden-Bell, D., 2002, astro-ph/0207064
\bibitem[\protect\citeauthoryear{Stackel}{1890}]{sta}
Stackel, P., 1890, Math.Ann., 35, 91
\bibitem[\protect\citeauthoryear{Whittaker}{1904}]{whi}
Whittaker, E.T., 1904, Analytical Mechanics pp 432, Cambridge 1959,
Edn.
\end{itemize}
\label{lastpage}
\end{document}